# Title

The carbon holdings of northern Ecuador's mangrove forests.

# Authors

Stuart Hamilton

John Lovette

Mercy Borbor

Marco Millones

# Abstract

Within a GIS environment, we combine field measures of mangrove diameter, mangrove species distribution, and mangrove density with remotely sensed measures of mangrove location and mangrove canopy cover to estimate the mangrove carbon holdings of northern Ecuador. We find that the four northern estuaries of Ecuador contain approximately 7,742,999 t (± 15.47 percent) of standing carbon. Of particular high carbon holdings are the *Rhizophora mangle* dominated mangrove stands found in-and-around the Cayapas-Mataje Ecological Reserve in northern Esmeraldas Province, Ecuador and certain stands of *Rhizophora mangle* in-and-around the Isla Corazón y Fragata Wildlife Refuge in central Manabí Province, Ecuador. Our field driven mangrove carbon estimate is higher than all but one of the comparison models evaluated. We find that basic latitudinal mangrove carbon models performed at least as well, if not better, than the more complex species based allometric models in predicting standing carbon levels. In addition, we find that improved results occur when multiple models are combined as opposed to relying any one single model for mangrove carbon estimates. The high level of carbon contained in these mangrove forests, combined with the future atmospheric carbon sequestration potential they offer, makes it a necessity that they are included in any future payment for ecosystem services strategy aimed at utilizing forest systems to reduce $CO_2$ emissions and mitigate predicted $CO_2$ driven temperature increases.

# Introduction

Mangrove forests provide numerous goods and services at the local, regional, national, and global levels (Ewel, Twilley, and Ong 1998; FAO Fisheries and Aquaculture Department 2004; Costanza et al. 2014). The most recent economic valuation of mangrove forests and the tidal swamps they inhabit is approximately $194,000 per hectare per year making them one of world's most economically productive ecosystems (Costanza et al. 2014). For example, at the national scale mangroves have been shown to sustain half of Malaysia's annual fish catch by providing habitat support, larval retention, and trophic supply (Chong 2007). Beyond fisheries support, mangroves provide numerous other goods and services. Mangroves directly support at least sixteen commercial food products, at least ten differing wood products, and provide at least eight environmentally important mitigation functions (FAO Fisheries and Aquaculture Department 2004; Hamilton and Collins 2013). Indeed, at the local level, coastal communities' livelihoods and food security are often intimately intertwined with the productivity of their local mangrove ecosystem which they exploit for cooking fuel, food, and timber (Conchedda, Lambin, and Mayaux 2011; Ocampo-Thomason 2006). Although mangroves provide numerous economic and environmental benefits, it is in the realm of global climate change science that mangrove forests likely provide their most important environmental and economic function.

Mangrove forests contain some of the highest global forest carbon stocks per unit area of any forest type (Intergovernmental Panel on Climate Change 2007; Alongi 2012; Donato et al. 2011). Current methods utilized to estimate global mangrove biomass, global mangrove carbon density, or global mangrove carbon emissions fall into two general categories. The first category are field based methods that establish mangrove biomass using measures such as tree height, stem density, tree diameter at breast height (DBH), and tree species. These field based studies most often sample in the species-rich mangrove region of the Indo-West Pacific (IWP) and then apply their findings across all mangroves globally. The second grouping is based on latitudinal estimates of mangrove carbon, usually based themselves on a synthesis of field studies, which are then also applied globally.

The latitudinal location of mangroves is important as mangrove biomass, and hence mangrove carbon density, are demonstrated to be inversely related to latitude (Twilley, Chen, and Hargis 1992; Saenger and Snedaker 1993) (Figure 1). That is, mangroves closer to the equator typically contain more biomass, and hence more carbon, than those farther away. The two primary drivers of this phenomenon are the mangrove responses to cooler water temperature and to lower levels of insolation. It is demonstrated that lower water temperatures result in reduced mangrove tree height and density (Lugo and Patterson-Zucca 1977) and that mangroves farther from the tropics are exposed to lower levels of insolation which inhibits growth when compared to tropical locations (Twilley, Chen, and Hargis 1992; Spalding, Blasco, and Field 1997).

**Figure 1. Mangrove carbon density in relation to latitude**

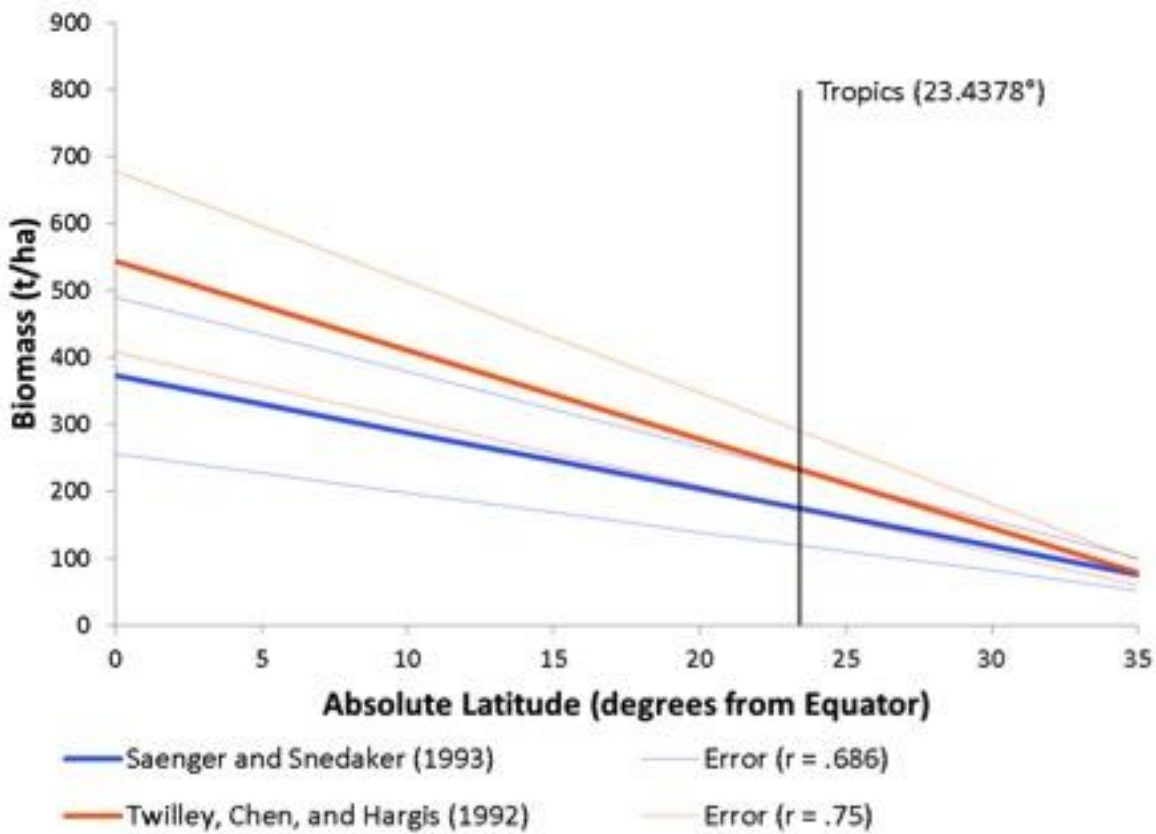

The x-axis represents absolute latitude moving away from the equator north or south and the y-axis represents the living mangrove biomass estimate in tonnes of a complete one hectare of mangrove at each degree of longitude.

Mangrove biomass and carbon density are not purely a function of their latitudinal location but also vary across mangrove species, which themselves show substantial regional variability. For example, the IWP region boasts the largest regional diversity of mangrove, with as many as 47 species identified as mangrove present in the coral triangle and 55 of the global total of 73 species present in Indonesia alone (Spalding, Kainuma, and Collins 2010). The general longitudinal trend is a decrease in mangrove diversity as you move away from the IWP in any direction with fewer than 10 species present throughout the Atlantic, Caribbean, and Eastern Pacific (ACEP) region. This facet of mangrove sorting is referred to as the mangrove anomaly and has been the subject of extensive research. The current paradigm is known as the vicariance hypothesis and "asserts that mangrove taxa evolved around the Tethys Sea during the Late Cretaceous, and regional species diversity resulted from *in situ* diversification after continental drift" (Ellison, Farnsworth, and Merkt 1999) pointing to a non-IWP point of origin and dispersal scenario. Ellison, Farnsworth, and Merkt (1999) additionally note a trifecta of independent regional diversification of mangrove species in South-east Asia, the Indian Ocean, and the ACEP region.

Within the field measure grouping of mangrove carbon estimates, Donato et al. (2011) sampled twenty-five mangrove locations in the IWP and calculated whole system carbon storage. They find that global mangrove forests contain on average 1023 t C ha$^{-1}$ ± 88 and of this 159 t C ha$^{-1}$, or 15.5 percent, is above ground living mangrove carbon. They then extrapolate these findings globally to calculate that mangrove clearing likely releases between 20,000,000 and 120,000,000 t C yr$^{-1}$. Unfortunately, this analysis, although reporting global figures, has no mangrove samples taken in the less species-rich ACEP region.

Regional mangrove field carbon estimates extrapolated globally should be treated with caution. This is particularly true if omitted sample regions are dominated by one or two species of mangrove that have carbon densities at the high end or low end of the carbon range when compared to carbon values found among mangroves in the species-rich IWP. That is, predicting mangrove carbon for homogeneous mangrove locations from heterogeneous mangrove locations when the homogeneous locations contain only species that are at the high-end or low end of the carbon holding range could results in unsatisfactory predictions as you move away from the IWP.

Within the latitudinal grouping of mangrove carbon estimates, Siikamäki, Sanchirico, and Jardine (2012) aggregate global mangrove cover from the Giri et al. (2011) remotely sensed mangrove presence data. From this remotely sensed mangrove cover data, they create a coarse 9-$km^2$ global mangrove grid that depicts mangrove presence or absence. They populate this grid with biomass values utilizing the latitudinal aboveground biomass function as demonstrated in Figure 1 (Twilley, Chen, and Hargis 1992). They then combine this information with the information from Donato et al. (2011), using 25 IWP samples for belowground biomass estimates. Mangrove soil carbon is estimated from a synthesis of literature. They find that mangrove forests globally contain on average 466.5 t C $ha^{-1}$ and place the global carbon stock at 16,500,000,000 t C. Of this 466.5 t C $ha^{-1}$, living biomass constitutes 147.5 t C $ha^{-1}$ or 31 percent of the total. Although the methodology has merit, the sampling procedure is again heavily influenced by IWP mangrove sample that may not be representative of the global mangrove species distribution, is at a coarse 9 $km^2$ resolution, and relies on latitude adjustments that have a high level of variance between models at the tropics (Figure 1) as well as within each model. For example, the Siikamäki, Sanchirico, and Jardine (2012) above ground carbon latitudinal model used only explains 75 percent of the variability in aboveground carbon at differing latitudes (Twilley, Chen, and Hargis 1992).

Limited research into mangrove carbon stocks exists for South America. Within Brazil, the aboveground living biomass carbon stock of mangrove has been shown to average 61.33 t C

ha$^{-1}$, with fringe mangroves averaging 90 t C ha$^{-1}$, basin mangroves 59 t C ha$^{-1}$ and transition mangroves 25 t C ha$^{-1}$ (Estrada et al. 2014). This approach utilized allometric equations and field measurements to calculate forest carbon. Utilizing similar allometric methods, total aboveground biomass for mangroves in French Guiana were measured at between 31 t ha$^{-1}$ for pioneer trees to 315 t ha$^{-1}$ in mature stands (Fromard et al. 1998), equating to approximately 14 t C ha$^{-1}$ and 146 t C ha$^{-1}$, respectively.

A paucity of research exists that examines the mangrove biomass or mangrove carbon questions within Ecuador. For example, none of the major mangrove aboveground or belowground global mangrove carbon estimates contains a single sample from Ecuador and neither do the major global mangrove carbon soil estimates. Within Ecuador, mangrove forest carbon was modeled using a combination of latitudinal non-species models and species specific allometric equations by Hamilton and Lovette (2015). They calculated combined living aboveground and belowground living mangrove carbon at 9,940,912 t C, or on average 262 ± 70 t C ha$^{-1}$.

In summary, Ecuadorian mangrove forests may not be accurately represented in the current mangrove carbon literature for two major reasons. Firstly, carbon estimates derived from field sampling may be unreliable when applied with field measures taken from Ecuador or even from neighboring countries. Most samples used to build biomass estimates are taken from the mangrove species-rich area of the IWP, whereas Ecuador is species limited with only one or two mangrove species dominating the entire mangrove area (Arriaga, Montaño, and Vásconez 1999; Madsen, Mix, and Balslev 2001; Spalding, Blasco, and Field 1997; Spalding, Kainuma, and Collins 2010) and the *Rhizophora mangle* that dominates in Ecuador is among the highest biomass of all mangrove species. Additionally, the current paradigm relating to mangrove evolution being a regional phenomenon combined with the substantial difference of biomass between differing mangroves makes the extrapolation of regional mangrove surveys to global values problematic. It is not only the species approach that may misrepresent Ecuadorian mangrove carbon but also potentially the latitudinal approach. Indeed, it is known that due to Ecuador's equatorial location, the forests along the country's coast receive some of the highest insolation of all mangrove forests and boast some of the largest trees in the world (Spalding, Blasco, and Field 1997; Spalding, Kainuma, and Collins 2010).

In this article, we overcome the identified research gap by estimating the carbon stock of Ecuador's equatorial mangrove forests from data collected within the estuaries of northern Ecuador. We calculate these carbon values using geographic information system (GIS), remote sensing, and estuarine specific field measurements of mangrove DBH, mangrove species, and stem density, which we then convert into carbon estimates using species-specific allometric equations from similar ACEP studies. We then compare our results to other estimates of mangrove carbon for this region extracted from global species models, latitudinal estimates, and Ecuador specific estimates. Finally, we discuss the implications of this research to wider global mangrove carbon stock including the implications of these findings of utilizing mangrove carbon to mitigate atmospheric $CO_2$.

## Study Area

The study area consists of the four major northern estuaries of Ecuador with latitudes ranging from 0.70° S to 1.46° N (Figure 2). Ecuador was selected for analysis due to its long history and availability of high-resolution spatiotemporal data for each estuary (Hamilton and Stankwitz 2012; Centro De Levantamientos Integrados De Recursos Naturales Por Sensores Remotos 2007; Hamilton and Lovette 2015), pre-established mangrove surveys (Centro De Levantamientos Integrados De Recursos Naturales Por Sensores Remotos 2007; Hamilton and Lovette 2015), and participation in payment for performance carbon programs (de Koning et al. 2011; Naughton-Treves and Day 2012; Holland et al. 2014; Hamilton and Stankwitz 2012). The study area encompasses 96,044 ha with a combined mangrove area of 37,953 ha as of 2011.

**Figure 2. Study Sites**

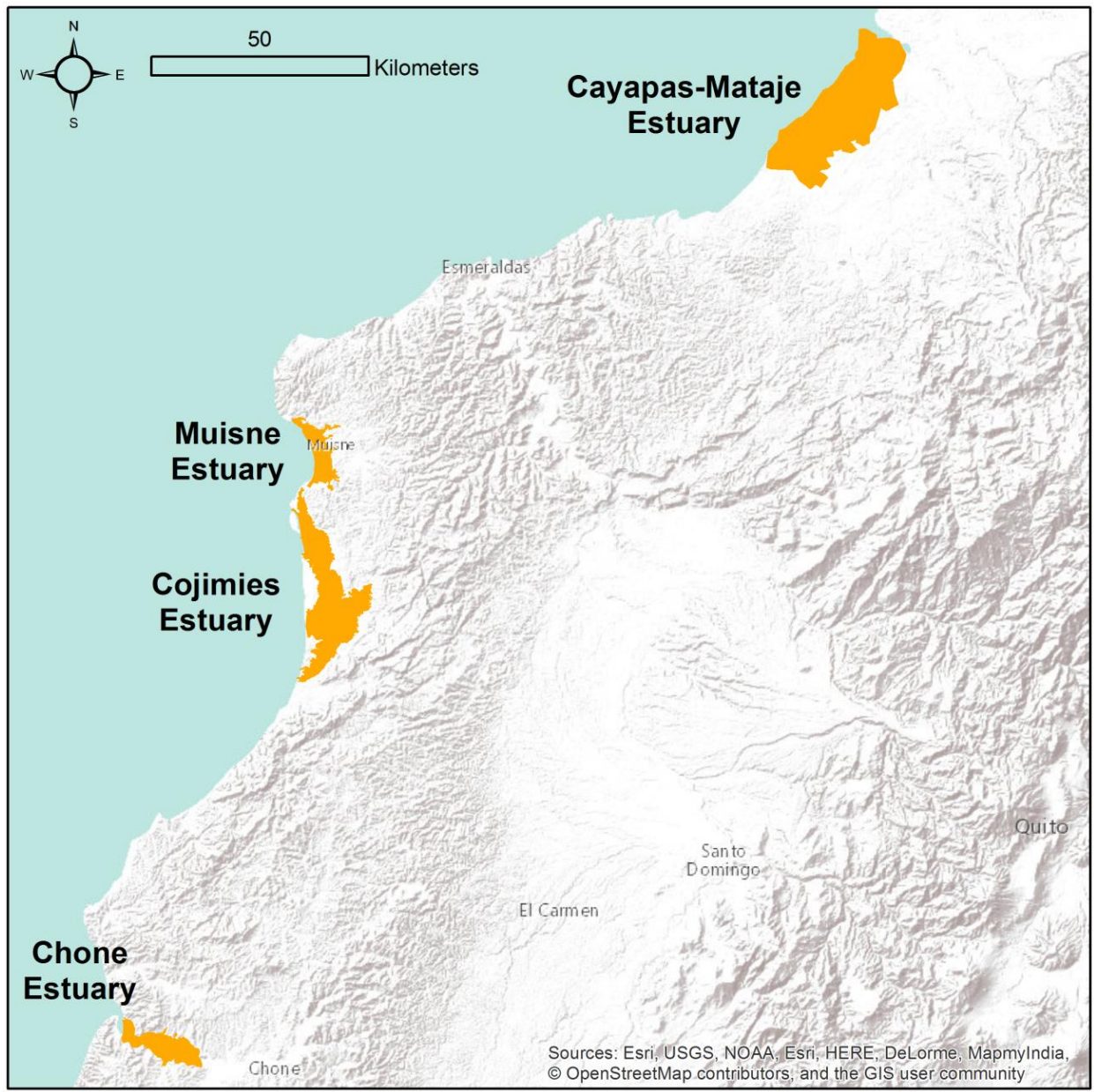

The study estuaries from north to south are (1) Cayapas-Mataje, (2) Muisné, (3) Cojimíes, and (4) Chone.

From north to south, the four estuaries are: (i) Cayapas-Mataje, located wholly within Esmeraldas province along the Colombian border, in-and-around the town of San Lorenzo; (ii)

Muisné, located wholly within Esmeraldas province near the town of the same name; (iii) Cojimíes, located on the border between Esmeraldas and Manabí north of the city of Pedernales; and (iv) Chone estuary, located wholly within Manabí province in-and-around the city of Bahia de Caráquez. These four estuaries are estimated to contain over 95 percent of the historic mangrove habitat in Ecuador's northern coastal provinces (Hamilton and Collins 2013).

Hamilton and Collins (2013) provide a thorough review of the protected status, land use dynamics, and local livelihoods exploitation of the mangrove forests of this region, while Hamilton and Stankwitz (2012) provide detailed information on the historic land use and land cover in each of the estuaries. Local residents anecdotally state that the mangroves of this region were substantially cleared across all four estuaries during the mid-twentieth century. This clearing is reflected in the academic literature as a government sponsored industrial program focused on exploiting the mangroves of this region for tannin (Labastida 1995; Ocampo-Thomason 2006; Snedaker 1986; Spalding, Blasco, and Field 1997; Hamilton 2012) that existed from the early 1950s until a collapse in tannin prices in the late 1960s.

By the late 1960s, as the tannin industry was collapsing, shrimp aquaculture arrived in Chone estuary and advanced north through Cojimíes and Muisné estuaries in the 1970s, eventually arriving in Cayapas-Mataje by the early 1980s (Hamilton 2012). The loss of mangroves in all estuaries, aside from Cayapas-Mataje, was dramatic during the period of aquaculture expansion, with 83 percent of all mangroves lost in the remaining three northern estuaries (Hamilton 2011). Of these losses, the majority of mangroves were directly displaced by shrimp aquaculture ponds (Hamilton and Lovette 2015; Hamilton 2013, 2012). Cayapas-Mataje mangroves appeared to have remained mostly undisturbed during this aquaculture-driven period of mangrove loss (Hamilton and Lovette 2015; Hamilton 2013, 2012). The post-2000 mangrove cover in all estuaries has remained essentially stable to present (Hamilton and Casey 2016), with three of the four estuaries actually having limited but measurable amounts of mangrove afforestation post-2000 (Hamilton and Lovette 2015; Hamilton 2011).

Due to recent conservation efforts, all mangrove stands in Ecuador have come under federal protection under Ecuadorian decree 001-DE-052-A-DE of 2013. Before 2013, each estuary's mangrove stands had differing levels of protection originating at different times. Almost all of the mangroves around Cayapas-Mataje fall in the Ecological Mangrove Reserve of Cayapas Mataje (REMACAM) which is an original Ramsar site (Wetlands International 2004) with the federal government, via the Ministry of the Environment, owning and overseeing the protection of the mangroves since at least 1995. From 2003 onwards, approximately 1 percent of the mangroves in Muisné are preserved through the privately operated Mangroves of the Muisné River Estuary Refuge (Resolution 047, 03-2003). However, much of the rest of the estuary appears without protection (Hamilton and Collins 2013). There appears to be very little, if any, government support or protection within the Cojimíes estuary (Herrera and Elao 2007; Crawford 2010). Voluntary local protection dominates in the Chone estuary, with local entities bringing much of the estuary under a special management since 1988 in hopes of improving the health of the estuary and its surrounding area (Arriaga, Montaño, and Vásconez 1999). Starting in 2002, the Chone estuary also had a small portion of the estuary protected as the Corazón and Frigatas Islands Wildlife Reserve (Registro Oficial No 733). Although public protection is limited, it is noted that community involvement through *custodias* and *concesiones* programs have been vital to the protection and conservation of many of Ecuador's mangrove forests (Ocampo-Thomason 2006).

## Materials and Methods

**Land Cover**

We utilize a subset of a previously created 1 ha estuarine land cover grid for all of Ecuador (Hamilton and Lovette 2015) to establish the 2011 location and amount of mangrove in each of the northern estuaries (Supp. Material, Figure 1). Within this grid are 96,044 1 ha cells, with each cell containing a land cover classification. Out of the 96,044 cells in the database, 49,728 cells contain some level of mangrove. The land cover classification process utilized by Hamilton and Lovette (2015) to obtain the mangrove cover across the estuaries and within each cell was an IsoData driven unsupervised classification process combined with field verification (Supp. Material, Figure 2). Input instruments were the Rapid Eye 5m resolution orthorectified system that captures spectral information between 440 nm and 850 nm and the ASTER 15m resolution system that captures spectral information between 520 and 850 nm. The resultant product is a 1 ha grid comprised of nested 5 m mangrove presence or absence cells derived from Rapid Eye imagery or 15 m cells obtained from ASTER imagery. Within the classification scheme utilized, the mangrove class contains percentage mangrove cover within each 1 ha cell and not merely mangrove presence or absence data. For example, a value of one within a mangrove cell indicates the entire cell is mangrove; a value of 0.1 indicates that 10 percent of the cell is mangrove, and a value of .5 indicates half of the cell is mangrove.

**Sampling**

Across all estuaries, each cell depicted as containing mangrove was extracted from the Hamilton and Lovette (2015) estuarine land cover database (Supp. Material, Figure 1). Mangrove pixels were then converted into mangrove stands by manual delineating areas of continuous mangrove forest cover from historic topographic maps and aerial photography (Supp. Material, Figure 2). This is particularly suitable as mangrove forests stands most often consist of a homogenous collection of single species trees (Hogarth 1999, 2007; Tomlinson 1986). Across all estuaries, potential sampling points were assigned to random locations at a point density of 1 point per km$^2$ (Supp. Material, Figure 2). The sampling process can be viewed as systematic at the estuarine scale with the estuary broken into logical stands, and as random at the mangrove stand level with points randomly assigned for field survey within the mangrove stands. Sampling was conducted prior to field data collection.

**Field Data**

The randomly assigned points within mangrove stands provided the suite of potential sites for field measurement of mangrove species, mangrove count, mangrove DBH, and mangrove height. One point that fell in an isolated stand northeast of the village of Bourbón, Esmeraldas was inaccessible due to being on private property and hence an alternate stand near to this point was surveyed on public property. Fieldwork was conducted during March 2013 and July 2013 under permit number 008-AT-DPAM-MAE granted by the Ministerio de Ambiente, Portoviejo, Ecuador. Within the field, mangrove carbon measurement protocols were adapted from the Center for International Forestry Research (CIFOR) standard (Kauffman and Donato 2012), DBH was obtained using the highest prop root as the starting point for the for *Rhizophora* *mangle* (Figure 3) and the traditional DBH (1.37 m above ground level) for other mangrove species. As opposed to using CIFOR circular plots, we utilized 10 m² square plots to better nest within the remote sensing derived 1 ha grid. As our analysis is concerned with living carbon, non-standing dead wood was excluded from the analysis although it is noted that one or two site locations contained substantial amounts of downed deadwood likely far exceeding the standing biomass.

**Figure 3. Field data collection**

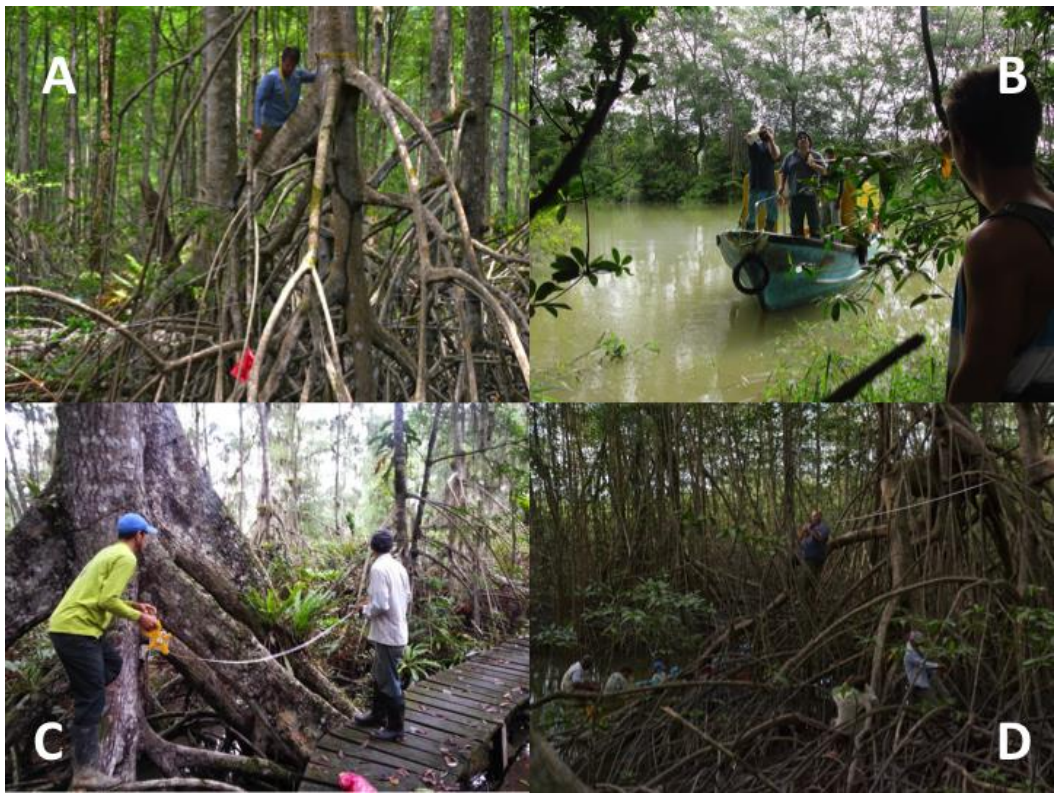

4A – Scaling *Rhizophora mangle* to obtain DBH measurement.
4B – Using a clinometer to obtain average stand height.
4C – Staking out a 10m by 10m survey grid among large *Rhizophora mangle* trees.
4D – The extensive root system for one *Rhizophora mangle* tree.

In eighteen mangrove stands across the four estuaries, twenty single-species mangrove surveys were performed. Survey teams traveled to these eighteen sites and centered plots as close to the random point as possible. GPS points were collected at the center of the plot and 10 m x 10 m survey areas were constructed around this point. Within each plot, all standing trees were tagged with their species, counted, and measured for DBH. We also estimated the tree stand height using a hand-held clinometer measuring the tree closest to the randomly selected GPS point at the center of each survey plot (Figure 3). Tree height was not utilized in any of our allometric equations but is included for later comparative analysis. In addition to noting species at the survey locations, 64 percent of total stands were field verified solely for species information. A nearest neighbor approach was then utilized to apply species information to remaining stands where no site visit occurred. In a vast majority of cases, almost all mangrove was designated with the *Rhizophora mangle* classification as this was the dominant species in each estuary.

**Biomass**

Numerous studies have formulated allometric equations relating DBH to biomass for the three mangrove species found in our studies sites (Soares and Schaeffer-Novelli 2005; Fromard et al. 1998; Imbert and Rollet 1989; Smith III and Whelan 2006) and it has been shown that allometric equations tend to be exhibit far less variation by site than they do by species (Ong 2002; Komiyama, Ong, and Poungparn 2008). For each survey plot, DBH was used to determine average tree biomass for each species present. By combining four allometric equations for *Rhizophora mangle* and three each for *Laguncularia racemosa* and *Avicennia germinans* (Table 1), species-derived biomass estimates were calculated for each plot. Utilizing the tree count within the plot, the stem density per hectare was also estimated. Combining these two measures, we determined per hectare aboveground biomass estimates for each survey plot.

**Table 1. Equations used as part of the allometric review to calculate tree and stand biomass.**

| Location | Lat | Lon | Species | Allometric Equation (individual tree (g)) | $R^2$ | DBH Range (cm) | Reference |
|---|---|---|---|---|---|---|---|
| **Brazil** | | | | | | | |
| *Bertioga* | 23°55' S | 46°20' W | *Rhizophora mangle* | $AGB = e^{(4.89219+2.61724*\ln(DBH))}$ | .991 | 1-20 | Soares and Schaeffer-Novelli (2005) |
| **French Guiana** | | | | | | | |
| *Mahury River* | 4°52' N | 52°19' W | *Rhizophora mangle* | $AGB = 128.2 * DBH^{2.6}$ | .92 | 1-32 | Fromard et al. (1998) |
| *to* | to | to | *Laguncularia racemosa* | $AGB = 102.3 * DBH^{2.5}$ | .97 | 1-10 | |
| *Counamama River* | 5°30' N | 53°10' W | *Avicennia germinans* | $AGB = 200.4 * DBH^{2.1}$ | .82 | 1-4 | |
| | | | *Avicennia germinans* | $AGB = 140.0 * DBH^{2.4}$ | .97 | 4-42 | |
| **Florida** | | | | | | | |
| *Everglades NP* | 25°08' N | 80°55' W | *Rhizophora mangle* | $AGB = 772.7 * DBH^{1.731}$ | .937 | 0.5-20.0 | Smith III and Whelan (2006) |
| | to | | *Laguncularia racemosa* | $AGB = 362.2 * DBH^{1.930}$ | .977 | 0.5-18.0 | |
| *Everglades NP* | 25°30' N | 81°12' W | *Avicennia germinans* | $AGB = 402.7 * DBH^{1.934}$ | .951 | 0.7-21.5 | |
| **Guadeloupe** | | | | | | | |
| *Grand Cul-de-Sac* | | | *Rhizophora mangle* | $AGB = 177.9 * DBH^{2.4176}$ | .98 | 6-23 | Imbert and Rollet (1989) |
| | 16°18' N | 61°32' W | *Laguncularia racemosa* | $AGB = 94.2 * DBH^{2.5367}$ | .99 | 7-26 | |
| | | | *Avicennia germinans* | $AGB = 208.8 * DBH^{2.239}$ | .99 | 6-41 | |

AGB: Aboveground Biomass, DBH: diameter at breast height (~1.37 m).

Belowground mangrove biomass was calculated for all surveyed areas as a function of the aboveground biomass. The literature does not contain the same wealth of belowground biomass allometric relationships as it does for aboveground biomass; however, many studies have synthesized global mangrove data to establish relationships between the aboveground and belowground biomass values. For example, Komiyama, Ong, and Poungparn (2008) found that the aboveground biomass to belowground biomass ratio averages *1:0.52*. We utilized this ratio to calculate approximate belowground biomass storage, and therefore total living biomass, for all mangrove trees and stands.

**Standing Carbon**

Finally, the total standing carbon stock was estimated for each surveyed mangrove stand. The conversion from total biomass to living carbon varies slightly throughout the literature, ranging from 1:0.45 - 0.50. Here we used a biomass to carbon ratio of *1:0.464* (Kauffman and Donato 2012; Kauffman et al. 2011). Using the allometrically calculated biomass and carbon values, biomass and carbon distribution by species was also interpolated across each estuary using a nearest neighbor function. That is, we utilized the species information and the species-specific allometric equations to calculate above ground biomass from our field measures of DBH, followed by conversion to total living biomass and carbon stock.

Species and living carbon distributions were calculated in the estuaries individually because of the differing influences of a variety of environmental variables in each study area. By combining the species distribution areas with the living carbon distributions for each species, spatially variable species-carbon values were assigned to mangrove stands throughout the estuaries. In the few mixed species stands present, percent dominance of the present species was observed during fieldwork and was used to calculate the biomass appropriately for that stand. For example, a stand may be any combination of *Rhizophora mangle*, *Laguncularia racemosa*, or *Avicennia germinans* as long as the values sum to 100 percent. Although mixed stands did exist within the estuaries they were few in number with most stands consisting of homogenous *Rhizophora mangle* of similar height and DBH.

A brief supplemental methods section provides a sample of the mangrove cover grid utilized (Supp. Material, Figure 1), a model that demonstrates how each cell in the grid was processed from mangrove cover to the final carbon estimate (Supp. Material, Figure 2) and a flowchart outlining the mangrove class field verification process (Supp. Material, Figure 3).

**Model Comparison**

To allow for comparison between differing mangrove carbon estimates we either obtained or built six differing mangrove carbon estimates in addition to the results provided. Three models were obtained directly from Hamilton and Lovette (2015) and used without any modification. Two of these three models are species driven modelled mangrove carbon estimates and one is an averaged mangrove carbon estimate that combines multiple input models including the two noted above. These three models have a spatial resolution of 1 ha and a treecover year of 2011, just as in this analysis. Indeed, we utilized the Hamilton and Lovette (2015) treecover estimate and their grid to facilitate direct comparison between results. These models are directly comparable to the results presented in this paper at all scales.

Two of the three other mangrove carbon estimates are species independent latitudinal models. The latitudinal models were built specifically for this paper and utilize mangrove equations from the mangrove literature to calculate mangrove carbon holdings. The equations are derived from Saenger and Snedaker (1993), Twilley, Chen, and Hargis (1992), and Komiyama,

Ong, and Poungparn (2008). Slight modifications of the equations were necessary to adjust for our 1 ha grid size. For these estimates we used the same input grid and mangrove cover year as used in this analysis and in the models provided by Hamilton and Lovette (2015). The mangrove cover year remains as 2011. These models are directly comparable to the results presented in this paper at all scales.

The final carbon model comes from Ruesch and Gibbs (2008) and is described as IPCC-compliant. This dataset was not adjusted or constructed for this analysis but used as provided. As opposed to all other methods, this dataset uses the Global Land Cover 2000 (GLC 2000) product to ascertain land cover in the estuary. Tree cover is treated as a binary variable in this grid with each grid having a singular landcover value without a percentage cover noted. GLC 2000 is built from the SPOT sensor in 2000 and is provided in a 1 km grid. The 1 km gird did require a slight shifting to allow our 1 ha to fully nest within it. This coarser grid results in only one data point for every one hundred data points for all other models. For these spatial and attribute mismatch reasons, the IPCC- compliant data was only analyzed at the entire estuarine level as opposed to the stand level or 1 ha grid level. Additionally, the IPCC-complaint land cover measure predates all the other models by 11-years. Due to the already noted actual increase in mangrove cover since 2000 across this region this should, all over things being equal, actually cause an overestimation of carbon by the IPCC-complaint when compared to this analysis and all other models utilized. For these reasons, caution is advised when comparing the IPCC data to our data at the sub-estuarine scale.

# Results

## Land Cover

Across the four estuaries, only the three primary species of *Rhizophora mangle*, *Avicennia germinans*, *and Laguncularia racemosa* were encountered during the field surveys. As suggested in the literature, *Rhizophora mangle* was dominant across all estuaries (Arriaga, Montaño, and Vásconez 1999; Madsen, Mix, and Balslev 2001; Spalding, Kainuma, and Collins 2010). Of the 85 plots pulled from the sample, 59 contained mangrove, with 80 percent of these being *Rhizophora mangle*. *Rhizophora mangle* occupied 98 percent, 71 percent, 70 percent, and 75 percent of the total mangrove areas in Cayapas-Mataje, Muisné, Cojimíes, and Chone estuaries respectively (Table 2). *Laguncularia racemosa* accounted for 15 percent of the estuarine mangrove cover across all estuaries with *Avicennia germinans* accounting for the remaining 5 percent (Table 2). *Laguncularia racemosa* coverage was relatively higher in Muisné and Cojimíes with coverage at 28 percent and 25 percent respectively, whereas Chone estuary had the largest percentage of *Avicennia germinans* at 10 percent coverage (Table 2). Additionally, all of the 59 plots depicted as mangrove were found to contain mangrove (Supp. Material, Figure 3).

**Table 2. Mangrove survey results by estuary and species**

|          | *Rhizophora mangle* | | | *Avicennia germinans* | | | *Laguncularia racemosa* | | |
|----------|-----|-------|--------|-----|------|--------|------|------|--------|
|          | %   | DBH   | Height | %   | DBH  | Height | %    | DBH  | Height |
| Cayapas  | 98  | 27.17 | 40.13  | 1.6 | --   | --     | 0.05 | --   | --     |
| Muisné   | 71  | 10.54 | 17.67  | 0.8 | --   | --     | 28   | 8.01 | 18.25  |
| Cojimíes | 70  | 15.67 | 20.61  | 5   | --   | --     | 25   | --   | --     |

| Chone   | 75 | 21.49 | 21.80 | 10  | 10.64 | 22.59 | 15 | 9.45 | 13.11 |
| Average | 80 | 18.73 | 27.79 | 4.8 | 10.64 | 22.59 | 15 | 8.49 | 16.53 |

Mean DBH and height by species and estuary. The average row reflects the count of samples at each estuary and therefore is not merely a mean of values in the table. DBH and height averages represent sampled stands; therefore, not all species are represented in each estuary.

**Field Data**

DBH values were highest in the pristine forests in Cayapas-Mataje with an average *Rhizophora mangle* stand DBH of 21.47 cm and an average *Rhizophora mangle* height of 40.13 m (Table 2). Additionally, it is likely that we have surveyed one the largest mangrove trees ever recorded in-and-around Cayapas-Mataje with a *Rhizophora mangle* DBH reading of 106 cm near 1.17° N, 79.08° W. Not only were exceptionally large *Rhizophora mangle* stands found in Cayapas-Mataje, we also found *Rhizophora mangle* stands with high DBH in the interior of Chone Estuary. The average DBH of *Rhizophora mangle* stands in Chone was 27.17 cm and the average height was 21.80 m (Table 2). These large DBH values are not isolated to single *Rhizophora mangle* stands but are found across large portions Cayapas-Mataje estuary and smaller portions the interior of Chone estuary. The *Rhizophora mangle a*verage DBH was 10.54 cm for Muisné and 15.67 cm for Cojimíes. *Rhizophora* height was relatively consistent across the estuaries of Muisné and Cojimíes, measuring 17.67 m and 20.61 m, respectively

**Standing Carbon**

The total carbon estimates, when corrected for species, DBH, tree cover, stem count and projected across each estuary; show a total living mangrove carbon stock of 7,742,999 t ± 15.47 percent (Table 3). Cayapas-Mataje exhibited the highest surveyed carbon stocks of the four estuaries, followed by Chone, Cojimíes, and Muisné (Table 3). This is not surprising when considering the DBH measurements noted above. Cayapas-Mataje also had the highest area-weighted mangrove carbon stock at 199 t C ha$^{-1}$. Chone was lower, measuring 125 t C ha$^{-1}$, and Muisné and Cojimíes were significantly smaller in their per hectare carbon stocks, measuring 34 t C ha$^{-1}$ and 35 t C ha$^{-1}$ respectively.

**Table 3. Survey carbon results and comparison to modeled values.**

| | Ecuador Field | GIS / RS Models | | | |
| --- | --- | --- | --- | --- | --- |
| | (1) RESULTS | (2) Hamilton and Lovette (2015) | | (3) IPCC | |
| Cayapas-Mataje | 6,961,915 | 6,243,737 | 90% | 4,253,300 | 61% |
| Muisné | 123,684 | 193,936 | 157% | 417,700 | 337% |
| Cojimíes | 273,349 | 471,478 | 172% | 1,921,500 | 702% |
| Chone | 384,051 | 205,416 | 53% | 277,100 | 72% |
| Total | 7,742,999 ± 15.47% | 7,114,567 | 92% | 6,869,600 | 89% |

| | Latitude Models | | | | Species Models | | | |
| --- | --- | --- | --- | --- | --- | --- | --- | --- |
| | (4) Twilley, Chen, and Hargis (1992) | | (5) Saenger and Snedaker (1993) and Komiyama, Ong, and Poungparn (2008) | | (6) Hamilton and Lovette (2015): Set species distribution | | (7) Hamilton and Lovette (2015): Modeled species distribution | |
| Cayapas-Mataje | 7,357,082 | 106% | 5,067,617 | 73% | 4,254,004 | 61% | 4,152,070 | 60% |
| Muisné | 524,942 | 424% | 361,009 | 292% | 302,554 | 244% | 286,412 | 232% |
| Cojimíes | 934,118 | 342% | 642,624 | 235% | 538,388 | 197% | 478,057 | 175% |
| Chone | 418,341 | 109% | 287,402 | 75% | 241,458 | 63% | 240,483 | 63% |
| Total | 9,234,483 | 119% | 6,358,652 | 82% | 5,336,404 | 69% | 5,157,022 | 74% |

Comparison of survey carbon results as a whole and by estuary in tonnes of carbon. These values are compared to modeled carbon estimates from Hamilton and Lovette (2015) and the IPCC global 1 km grid estimates (Ruesch and Gibbs, 2008). The percentage of (1) are shown for each of the additional modeled carbon estimates. The error ± in column 1 is the standard error of the mean.

**Model Comparison**

Of all models presented, the Hamilton and Lovette (2015) data synthesis mangrove carbon estimation method is in closest agreement with the mangrove carbon results generated by this analysis (Table 3). It represents an overall underestimation of mangrove carbon of only 8 percent across the entire region when compared to our findings and is within the margin of error of our result (Table 3, column 2). Although the overall mangrove carbon estimation produced by the Hamilton and Lovette (2015) data synthesis approach is in close agreement with the results presented in this analysis, substantial disagreements occur at the estuarine level. For example, Cayapas-Mataje was underestimated by only 10 percent whereas Chone was actually underestimated by 47 percent. On the other hand, Muisné and Cojimíes are actually overestimated by 72 percent and 57 percent respectively. The close agreement between the Hamilton and Lovette (2015) average results and our results is driven by the fact Cayapas-Mataje is magnified in its importance as it is larger than the other three study sites combined, and the mangrove carbon estimates produced for Cayapas-Mataje across both studies are in close approximation.

The two modeled mangrove carbon stock prediction methods, utilizing species information, from Hamilton and Lovette (2015) do not perform as well as their data synthesis approach and substantially under estimate carbon values, when compared to our findings (Table 3, column 6 and 7). Their species driven models predict mangrove carbon levels that are 31 percent and 26 percent lower than those generated in this analysis. As with their data synthesis approach, their species modeled findings approach shows substantial variance across the estuaries. For example, both Cayapas-Mataje and Chone are somewhat underestimated whereas Muisné and Cojimíes are substantially overestimated.

When comparing the IPCC complaint 1 km grid carbon estimates (Ruesch and Gibbs 2008) to the field-derived estimates generated by this analysis, the IPCC-complaint carbon stocks are 11 percent lower than our estimate, and the differences again show considerable variation across the differing estuaries (Table 3, column 3). The IPCC-complaint data predicts carbon stocks of 4,243,300 t, 417,700 t, 1,921,500 t, and 277,100 t for each of Cayapas-Mataje, Muisné, Cojimíes, and Chone respectively for the year 2000. These values are 61 percent, 337 percent, 702 percent, and 72 percent of the survey driven value sin 2011, with Cayapas-Mataje once again responsible for most of the underestimation. Due to the differing treecover dates used in the IPCC-compliant data and our analysis, and the fact the mangrove cover is either stable or increasing since 2000, it is likely that the underestimation from the IPCC-complaint is actually larger than reported.

The two latitudinal mangrove carbon models that do not account for species differences produced carbon estimates are 19 percent higher and 18 percent lower than the survey carried out in this analysis (Table 3, columns 4 and 5). The Twilley, Chen, and Hargis (1992) latitudinal estimates are the only method that overestimated carbon levels when compared against our measure (Table 3, column 4). The other latitudinal method derived from Saenger and Snedaker (1993) and Komiyama, Ong, and Poungparn (2008) is 18 percent lower than our current estimate (Table 3, Column 5). The Saenger and Snedaker (1993) latitudinal method combined with the (Komiyama, Ong, and Poungparn 2008) allometric equations followed the pattern of over-estimating mangrove carbon in the central estuaries of Muisné and Cojimíes and understanding mangrove in Cayapas-Mataje and Chone. The Twilley, Chen, and Hargis (1992) latitudinal model almost identically matched our estimations in both Cayapas-Mataje and Chone but again overestimated mangrove carbon in the central estuaries of Cojimíes and Muisné. Interestingly, the average of both latitudinal models would vary by less than 1 percent from the results of this analysis.

Full results from this study at the 1 ha grid level are provided in open GIS format in the supplemental material ([Data Link](Data Link)).

**Results breakdown**

To demonstrate the difference in carbon finding between this survey and earlier models we will utilize the *Rhizophora mangle* plot (10 m x 10 m) at 1.23708°N, 79.04455°W as an example. The DBH and stem density values for this plot (Table 4) were then applied to allometric equations for *Rhizophora mangle* as described in Table 1 to establish per-hectare aboveground and belowground biomass values for the stand and then converted to total stand carbon. Based on the surveyed species distribution across the estuary this value, along with the other carbon values calculated from stand surveys, was propagated to spatially contiguous 1 ha analysis grid cells. The grid cell encompassing this survey site was established to have a carbon stock of 152 t C ha$^{-1}$. The same grid cell has a predicted mangrove carbon stock ranging between 140 t C ha$^{-1}$ to 242 t C ha$^{-1}$ as calculated by two species and two latitudinal models presented in Hamilton and Lovette (2015). The IPCC-complaint grids estimate is 193 t C ha$^{-1}$.

**Table 4. Survey results for a *Rhizophora mangle* plot in the Cayapas-Mataje estuary**

| Survey Point | CM46 |
|---|---|
| Latitude | 1.23708 |
| Longitude | -79.04455 |
| Average DBH (cm) | 20.33 |
| Maximum DBH (cm) | 44.88 |
| Stand Height (m) | 35.6 |
| n ( >2 cm DBH) | 8 |
| Species | *Rhizophora mangle* |
| Density (stems / ha) | 800 |
| Carbon (tonnes / ha) | 152 |

**Figure 4. Walkthrough of carbon stock calculation in 1 ha grid cell containing survey plot in the Cayapas-Mataje estuary**

(1) Surveyed Carbon Stock

$$AGB = e^{(4.89219+2.61724*\ln(DBH))}$$
$$AGB = 128.2 * DBH^{2.6}$$
$$AGB = 772.7 * DBH^{1.731}$$
$$AGB = 177.9 * DBH^{2.4176}$$

| Average DBH (cm) | 10 x 10 plot (grams per tree) | | | | Survey Grid (per hectare) | |
| --- | --- | --- | --- | --- | --- | --- |
| | Aboveground Biomass | Belowground Biomass | Combined Living Biomass | Living Carbon | Stem Density | Living Carbon (tonnes/ha) |
| (surveyed) | (from Table 1) | AGB * 0.52 | AGB + BGB | CB * 0.464 | (surveyed) | (LC (tonnes) * Stem density) |
| 20.33 | 353,477 | 183,808 | 537,284 | 249,300 | 8000 | 1,994 |
| | 322,883 | 167,899 | 490,782 | 227,723 | | 1,822 |
| | 142,035 | 73,858 | 215,894 | 100,175 | | 801 |
| | 258,660 | 134,503 | 393,163 | 182,428 | | 1,459 |
| | | | | | Average = | 1,519 |

(4) Twilley, Chen, and Hargis (1992)
$$CC_i (t.ha^{-1}) = (M_{it} (.464 (543.27 - 13.269 |Lat|)))$$
$$CC_i (t.ha^{-1}) = (1 (.464 (543.27 - 13.269 |1.23708|)))$$
$$CC_{46} = 242 \text{ t C ha}^{-1}$$

(5) Saenger and Snedaker (1993) and Komiyama, Ong, and Poungparn (2008)
$$CC^*_i (t.ha^{-1}) = (M_{it} (.464 (373.273 - 8.486 |Lat|)))$$
$$CC^*_i (t.ha^{-1}) = (1 (.464 (373.273 - 8.486 |1.23708|)))$$
$$CC^*_{46} = 167 \text{ t C ha}^{-1}$$

(6) Hamilton and Lovette (2015): Set Species Distribution
$$CC'_i (t.ha^{-1}) = (M_{it} (.464 (0.90 * BM_R + 0.05 * BM_B + 0.05 * BM_W)))$$
$$CC'_i (t.ha^{-1}) = (1 (.464 (0.90 * 297.54 + 0.05 * 293.01 + 0.05 * 359.83)))$$
$$CC'_{46} = 140 \text{ t C ha}^{-1}$$

(7) Hamilton and Lovette (2015): Modeled Species Distribution
$$CC^{\wedge}_i (t.ha^{-1}) = (M_{it} (.464 (nM_{Rit} * BM_R + nM_{Bit} * BM_B + nM_{Wit} * BM_W)))$$
$$CC^{\wedge}_i (t.ha^{-1}) = (1 (.464 (.407 * 297.54 + .222 * 293.54 + .370 * 359.83)))$$
$$CC^{\wedge}_{46} = 148 \text{ t C ha}^{-1}$$

The carbon stock for the 1 ha grid cell containing the survey point at 1.23708° N, 79.04455° W was calculated using the allometric equations from Table 1 and the carbon stock equations from Hamilton and Lovette (2015). This grid cell was completely forested as of the 2014 survey, and the measured plot was covered by *Rhizophora mangle*. The numbers next to each equation set refer to the total carbon stock calculations in Table 3.

**Error and Uncertainty**

The methods developed in this paper to calculate living mangrove carbon rely on numerous data inputs with error and uncertainty involved in each step of the process. In addition, the datasets used for comparative analysis likely contain the same potential errors and uncertainty as the data generated. For example, the EOS utilized likely have instrument error and atmospheric induced variability, and numerous other opportunities exist for the introduction of error and uncertainty in remotely sensed data (Jensen 2005; Jensen 2007). Once the remote sensing data is obtained and processed, the mere transformation of the data into higher-level GIS products has the potential to introduce at least five differing sources of uncertainty (Gahegan and Ehlers 2000). To account for error and uncertainty in this analysis we describe the reported error for each input dataset as well as conducting sensitivity analysis of the carbon findings in the form of a Leave-one-out cross validation (LOOCV).

**Land cover**

Hamilton and Lovette (2015) provide error bars for their modeled calculation of mangrove carbon, reporting ±0.27 percent, but do not specify the confidence interval of these error bars. They do not provide an error metric for their mangrove delineation that was used in this analysis but this was overcome by utilization of the field verified mangrove species measures taken as part of this study. As part of the fieldwork for this paper, we verified 85 locations across all four estuaries to ascertain the land cover type, and if mangrove were present, we would additionally ascertain the species of mangrove. The species data can be collapsed into a single class of mangrove and used to ascertain the accuracy of the mangrove delineation. Although limited errors existed in the non-mangrove classifications, all 59 sites delineated as mangrove were found to be mangrove forest in the field and none of the 26 areas delineated by as non-mangrove was found to be mangrove (Supp. Material, Figure 3). This resulted in a 100 percent accuracy rate in the LULC mangrove classification test nullifying the use of map comparison statistics such as kappa that are often used to generate land cover uncertainty metrics.

Testing of the mangrove tree cover values was not possible beyond the verification of presence or absence of mangrove as no higher resolution or more authoritative source of mangrove treecover exists to allow for comparison. A recently released global mangrove database (Hamilton and Casey 2016) may offer the potential to construct such a test but was not available at the time of this analysis.

**Latitudinal model accuracy**

Saenger and Snedaker (1993), do not provide error or uncertainty estimates for their stated relationship between latitude AGMB. They do however provide an *r*-value of 0.68 when using latitude to predict AGMB and 0.81 when latitude and height are combined into a single AGMB predictor. Both of these values are reported as significant at the $p < 0.0001$ level. Similarly, Twilley, Chen, and Hargis (1992), do not provide error measurements or uncertainty estimates for their stated relationship between latitude and CMB but do provide an *r*-value of 0.75 when calculating the relationship between latitude and CMB. They do not indicate the level of significance of this *r*-value. Despite this fact, the Twilley, Chen, and Hargis (1992) measure is the most commonly used latitudinal biomass predictor in the wider mangrove carbon and biomass literature e.g. (Siikamäki, Sanchirico, and Jardine 2012; Rovai et al. 2016; Donato et al.

2011). Using the provided correlation coefficients from both of these papers, we provide potential uncertainty bars in the latitudinal estimates presented in Figure 1.

**Allometric and equation uncertainty**

Four *Rhizophora mangle,* three *Avicennia germinans*, and three *Laguncularia racemosa* allometric equations outputs are averaged to form the basis of the biomass estimated presented. Although averaged allometric outputs are utilized in this analysis substantial variability can occur between allometric equations within the same species. We provide estimates for 10 cm DBH and 20 cm DBH mangrove trees across all mangrove species, across all equations, and report both biomass and carbon in Supplementary Table 1. This allows readers to ascertain the range of potential biomass values in addition to those presented.

The first allometric equation utilized (Table 1) in this study is for *Rhizophora mangle* and was developed by Soares and Schaeffer-Novelli (2005) sampling seventy-nine trees in Southern Brazil. Multiple regression models were tested to determine the best predictors for each biomass component, as well as for the whole tree. The authors report their equations significance at $p$ <0.01 with an adjusted $R^2$ of 0.991, and adjusted standard error of the estimate of $SSE = 4161$. The second set of equations utilized (Table 1) are developed by Fromard et al. (1998) who provide allometric equations for all three relevant species in French Guiana. These equations were built to relate biomass to DBH based on its ease of measurement and the propensity of other structural relationships (e.g. to crown diameter) to fail, especially in *Avicennia germinans*. Fromard et al. (1998) report $R^2$ of >0.82 for each of their four relationships and report that all are significant at $p$ <0.0001. Smith III and Whelan (2006) developed biomass equations for the three species within the Everglades National Park in Florida, USA. They specifically sampled trees in good health and with relatively straight trunks. The authors predetermined the form of their equation based on similar work in other regions. All regression equations were significant at $p \leq 0.05$ and had a total explained variance of at least 0.937. Imbert and Rollet (1989) developed their allometric equations for the three primary species in Guadaloupe. Similar to Smith III and Whelan (2006), the authors predetermined the form of their regressions. They report $r$-values >0.98 for all species.

A unique and potentially large contribution to uncertainty and error in this study is that some of the individual *Rhizophora mangle* surveyed in this analysis in Cayapas-Mataje are far larger than the *Rhizophora mangle* used to generate the allometric equations that exist within the academic literature. This requires the allometric equations to apply outside of the range of data used to construct the equations. Such extrapolation has the potential to overestimate mangrove biomass if older larger trees do not contain carbon, or sequester $CO_2$, at the same level as the smaller mangrove trees used to develop the allometric equations. Fortunately , the most recent global DBH analysis of over 650,000 trees, across 403 species, and including species classified as mangrove, indicates that the rate of tree carbon accumulation actually increases continuously with tree size and tree age (Stephenson et al. 2014a). This would actually result in a limited underestimation of carbon in particularly large mangrove trees, as not only does the carbon level of trees continue to increase with age and size, but the actual rate of carbon accumulation increases over time and tree size (Stephenson et al. 2014a). A brief discussion of the biomass to conversion factor is included in the supplementary material.

**Cross validation**

As now higher-level dataset exists for map comparison, we used a cross validation approach to examine potential uncertainty in the carbon estimates at the one ha gird level based or our field sample findings. We use the LOOCV framework (Refaeilzadeh, Tang, and Liu 2009), to estimate one left out carbon observation using all remaining filed samples and repeat this process dropping each individual sample in an iterative manner. We use the same nearest neighbor allocation method utilized in the methodology to calculate dropped samples. We then construct a simple standard error (SE) measure as well as two goodness of fit measures using this framework. The two goodness of fit measures presented are the traditional root mean square error (RMSE) and the lesser used normalized median absolute deviation (NMAD) that is likely a more suitable measure based on our sample data. NMAD can be, "considered as an estimate for the standard deviation more resilient to outliers in the dataset" (Höhle and Höhle 2009) (p.400). This is the case in this analysis were one or two *Rhizophora Mangle* samples contain are far higher levels of carbon than almost all the other samples in addition to our sample size being relatively small.

Due to the species configuration of the estuaries and because of sample size limitations we use only *Rhizophora Mangle* samples in the cross validation. The (LOOCV) $\bar{x}$ = 122.59, *Md* = 102.07, *s* = 70.24, RMSE = 105.14, SE = 18.14, with an NMAD = of 37.48. The NMAD was then utilized as a substitute for the standard deviation (Höhle and Höhle 2009) to construct confidence intervals at the 95 percent levels. These are the confidence intervals reports in the results section and abstract. If users of the GIS data prefer to construct confidence intervals using the standard error or standard deviation the information is provide above. One additional find of the cross validation is that that each field data sample introduces important information that anchors the carbon finding geographically in manner that coarse grids or generalized spatial models may not.

## Discussion

Northern Ecuador's mangrove carbon holdings are estimated to be higher than all currently species-specific extrapolated field measurements of mangrove carbon, modeled IPCC-compliant based estimates of estuarine carbon, and one of the two latitudinal estimates of mangrove carbon (Table 3). The Hamilton and Lovette (2015) synthesis measure provides the closest estimate of mangrove carbon when compared to our study and an average differing by only 8 percent. Combining both of the latitudinal models into a single model would actually produce almost identical results to those presented, even outperforming the Hamilton and Lovette (2015) synthesis measure. The fact that the two most similar results are averaged models suggests that combing models may produce more representative mangrove carbon estimates than using any one single model. In addition, in this example simple latitudinal models that are easy to implement may be more accurate in predicting living carbon for mangrove forests than the more complex and difficult to implement mangrove species based allometric models.

Hamilton and Lovette (2015) modeled mangrove carbon for the same group of estuaries as presented in this analysis. Their species-specific mangrove carbon measures are approximately 30 percent lower than the results reported in this table (Table 3). Due to the similar structure of both analyses such differences must be caused be either; (i) the field-measured distribution of mangrove species as opposed to the modeled mangrove species, or (ii) the field measured mangrove physical characteristics as opposed to the values taken from the literature synthesis. The reason these two components can be extracted as causing the differences is that the remaining parameters such a calculated mangrove area, mangrove grid size, mangrove grid location, the temporal component of the mangrove measure, and the classification systems are essentially consistent across both analyses. Additionally, (i) mangrove species can be ruled out as a major factor contributing to the differences as one of the models presented contains a simulation were 90 percent of the northern estuaries contain mature *Rhizophora mangle* which is comparable to the species distribution measured in our study. Therefore, all that remains to explain the differences in the mangrove carbon estimates are the field obtained mangrove DBH or the field measured mangrove tree density utilized in this study.

    The primary driver of high carbon levels found in this analysis is the dominance large *Rhizophora mangle* trees found in both Cayapas-Mataje and Chone. *Rhizophora spp.* exhibit some of highest biomass values of any mangrove at a variety of latitudes (Ricklefs and Latham 1993; Saenger and Snedaker 1993; Komiyama, Ong, and Poungparn 2008). Therefore, not only is Ecuador dominated by *Rhizophora mangle,* which is known to have some of the highest biomass of any mangrove specie globally, but also the field measurements of *Rhizophora mangle* DBH and height recorded contain some of the highest values within this species. Additionally within Ecuador, only between 10 percent and 20 percent of lower biomass species exist at the estuarine level to counter the high biomass of *Rhizophora mangle* when combined into an aggregate measure such as regional or even national mangrove carbon estimates. The presence of the largest individual mangrove trees in the world in this region has been known for some time (Spalding, Blasco, and Field 1997; Spalding, Kainuma, and Collins 2010) with reported heights of over 60 m (Ochoa 1997) but the prevalence of such stands is likely revealed for the first time

in this analysis. Not only did Cayapas-Mataje contain many stands of the largest *Rhizophora mangle*, or any other mangrove species, ever recorded but we also found large high biomass *Rhizophora mangle* stands in the interior of Chone Estuary.

Within Cayapas-Mataje and Chone, our findings for *Rhizophora mangle* DBH and height are typically above the values reported in much of the mangrove literature. For example, Pool, Snedaker, and Lugo (1977) surveyed *Rhizophora mangle* mangrove stands in Costa Rica. The survey area, dominated by *Rhizophora mangle* as in our estuaries, exhibited a relatively equal DBH distribution of around 10 cm and the trees only averaged about 10 m in height. Approximately the same values were reported when *Rhizophora mangle* was measured in Florida (Smith III and Whelan 2006). *Rhizophora mangle* has also been studied along the Atlantic coast of South America where maximum DBH measurements reach approximately 20 cm in mature stands (Fromard et al. 1998). Indeed, the average DBH of the *Rhizophora mangle* in Cayapas Mataje and in parts of Chone are much closer to the maximum values found in other Ecuadorian, Latin American, and global studies. Although it appears most trees continue to add biomass and sequester carbon at steady or increasing rates as they age (Stephenson et al. 2014b), the results presented in this paper would benefit from allometric equations developed in this region for the *Rhizophora mangle* stands that fall outside the DBH range on which the non-Ecuadorian preexisting allometric equations are built. Conversely, this would require removing a large number of healthy, mature *Rhizophora mangle* trees from a region that has only recently began to recover from widespread mangrove deforestation.

As of 2014, Ecuadorian mangroves forests are estimated to contain only approximately 1.16 percent ± 0.01 percent of all global mangrove forests (Hamilton and Casey 2016). Therefore, although insignificant as a portion of total global mangrove forest carbon, it is likely that these forests contain some of the highest living carbon storage values per hectare of any mangrove forests globally in additional to having one of the highest economic values of any natural ecosystem (Costanza et al. 2014; Costanza et al. 1997). Indeed, as mangrove forests are now accepted as containing some of the highest carbon values per hectare of any forest type, the mangrove forests of northern Ecuador likely contain some of the highest carbon sinks per hectare of not just mangrove but of any forest type globally. This is additionally relevant to the climate change mitigation role of these mangrove forests. In addition to their substantial carbon holdings, mangroves are already noted as being more efficient at sequestering $CO_2$ than other tropical forests (Laffoley and Grimsditch 2009), burying the sequestered carbon underground via an efficient year-round littering system (Chmura et al. 2003). Indeed, it is not only the current and future role of these Ecuadorian mangroves in sequestering atmospheric $CO_2$ that is likely enhanced by this research but also the levels of carbon that could potentially be released into the atmosphere if these mangroves are cleared. This is particularly relevant as Ecuador has historically been at the forefront of mangrove to aquaculture conversion and has suffered some of the highest rates of mangrove loss globally since 1970 (Hamilton 2013).

The implications of these findings in Ecuador to payment for ecosystem services programs such as the UN flagship Reducing Emissions from Deforestation and Forest Degradation (REDD+) program are likely regionally significant. If the northern mangrove forests of Ecuador do indeed contain higher carbon levels and atmospheric carbon sequestration potential than modeled by other methods, then the inclusion of mangrove in such programs as REDD+ becomes even more critical. Considering these new findings, mangrove forest conservation in northern Ecuador may provide a cost effective means of contributing to the stabilization of atmospheric carbon levels while providing all the other ecosystem goods and services that mangrove forests are known to provide.

# References

Alongi, D. M. 2012. Carbon Sequestration in Mangrove Forests. *Carbon Management* 3 (3):313-322.

Arriaga, L., M. Montaño, and J. Vásconez. 1999. Integrated Management Perspectives of the Bahia de Caráquez Zone and Chone River Estuary, Ecuador. *Ocean & Coastal Management* 42 (2-4):229-241.

Centro De Levantamientos Integrados De Recursos Naturales Por Sensores Remotos. 2007. Actualizacion Del Estudio Multitemporal de Manglares, Camaroneras Y Areas Salinas En Las Costa Continental Ecuatoriana Al Ano 2006, ed. O. P. Ramiro, 1-11. Quito, Ecuador: Centro De Levantamientos Integrados De Recursos Naturales Por Sensores Remotos.

Chmura, G. L., S. C. Anisfeld, D. R. Cahoon, and J. C. Lynch. 2003. Global Carbon Sequestration in Tidal, Saline Wetland Soils. *Global Biogeochemical Cycles* 17 (4):1111.

Chong, V. C. 2007. Mangroves-Fisheries Linkages in the Malaysian Perspective. *Bulletin of Marine Science* 80 (3):755-772.

Conchedda, G., E. F. Lambin, and P. Mayaux. 2011. Between Land and Sea: Livelihoods and Environmental Changes in Mangrove Ecosystems of Senegal. *Annals of the Association of American Geographers* 101 (6):1259-1284.

Costanza, R., R. d'Arge, R. de Groot, S. Farberk, M. Grasso, B. Hannon, K. Limburg, S. Naeem, R. V. O'Neill, J. Paruelo, R. G. Raskin, P. Sutton, and M. van den Belt. 1997. The Value of the World's Ecosystem Services and Natural Capital. *Nature* 387:253-260.

Costanza, R., R. de Groot, P. Sutton, S. van der Ploeg, S. J. Anderson, I. Kubiszewski, S. Farber, and R. K. Turner. 2014. Changes in the global value of ecosystem services. *Global Environmental Change* 26:152-158.

Crawford, B. 2010. *The Cojimies Estuary in Ecuador*. Coastal Resources Center: Graduate School of Oceanography. University of Rhode Island. 2010 [cited 12/1/2010 2010]. Available from http://www.crc.uri.edu/download/Ecuador_Cojimies_Program_Brief.pdf.

de Koning, F., M. Aguinaga, M. Bravo, M. Chiu, M. Lascano, T. Lozada, and L. Suarez. 2011. Bridging the Gap between Forest Conservation and Poverty Alleviation: The Ecuadorian Socio Bosque Program. *Environmental Science & Policy* 14 (5):531-542.

Donato, D. C., J. B. Kauffman, D. Murdiyarso, S. Kurnianto, M. Stidham, and M. Kanninen. 2011. Mangroves among the Most Carbon-Rich Forests in the Tropics. *Nature Geoscience* 4 (5):293-297.

Ellison, A. M., E. J. Farnsworth, and R. E. Merkt. 1999. Origins of Mangrove Ecosystems and the Mangrove Biodiversity Anomaly. *Global Ecology and Biogeography* 8 (2):95-115.

Estrada, G. C. D., M. L. G. Soares, V. Fernadez, and P. M. M. de Almeida. 2014. The economic evaluation of carbon storage and sequestration as ecosystem services of mangroves: a case study from southeastern Brazil. *International Journal of Biodiversity Science, Ecosystem Services & Management* (ahead-of-print):1-7.

Ewel, K. C., R. Twilley, and J. E. Ong. 1998. Different Kinds of Mangrove Forests Provide Different Goods and Services. *Global Ecology and Biogeography Letters* 7:83-94.

FAO Fisheries and Aquaculture Department. 2004. Mangrove Forest Management Guidelines. In *FAO Forestry Paper 117*, ed. F. F. a. A. Department, 319. Rome: FAO Fisheries and Aquaculture Department.

Fromard, F., H. Puig, E. Mougin, G. Marty, J. L. Betoulle, and L. Cadamuro. 1998. Structure, Above-Ground Biomass and Dynamics of Mangrove Ecosystems: New Data from French Guiana. *Oecologia* 115 (1-2):39-53.

Gahegan, M., and M. Ehlers. 2000. A framework for the modelling of uncertainty between remote sensing and geographic information systems. *ISPRS Journal of Photogrammetry and Remote Sensing* 55 (3):176-188.

Giri, C., E. Ochieng, L. L. Tieszen, Z. Zhu, A. Singh, T. Loveland, J. Masek, and N. Duke. 2011. Status and Distribution of Mangrove Forests of the World Using Earth Observation Satellite Data. *Global Ecology and Biogeography* 20 (1):154-159.

Hamilton, S. E. 2011. Quantifying mangrove deforestation in Ecuador's northern estuaries since the advent of commercial aquaculture. *ISME/GLOMIS Electronic Journal* 9 (1):1-3.

———. 2012. *The Impact of Shrimp Farming on Mangrove Ecosystems and Local Livelihoods along the Pacific Coast of Ecuador*. 1st ed. Hattiesburg, Mississippi: ProQuest, UMI Dissertation Publishing (October 18, 2012).

———. 2013. Assessing the Role of Commercial Aquaculture in Displacing Mangrove Forest. *Bulletin of Marine Science* 89 (2):585-601.

Hamilton, S. E., and D. Casey. 2016. Creation of a high spatio-temporal resolution global database of continuous mangrove forest cover for the 21st century (CGMFC-21). *Global Ecology and Biogeography* 25 (6):729-738.

Hamilton, S. E., and S. Collins. 2013. Las Respuestas a Los Medios De Subsistencia Deforestación De Los Manglares en Las Provincias Del Norte De Ecuador. *Bosque* 34 (2):143-153.

Hamilton, S. E., and J. Lovette. 2015. Ecuador's mangrove forest carbon stocks: a spatiotemporal analysis of living carbon holdings and their depletion since the advent of commercial aquaculture. *PLoS ONE* 10 (3):e0118880.

Hamilton, S. E., and C. Stankwitz. 2012. Examining the relationship between international aid and mangrove deforestation in coastal Ecuador from 1970 to 2006. *Journal of Land Use Science* 7 (2):177-202.

Herrera, M., and R. Elao. 2007. Analisis de Amenazas a La Biodiversidad en el Estuario De Cojimies (Ecuador) : Identificacion de Amenazas y Oportunidades Para El Manejo del Estuario de Cojimies y Su Area de Influencia Dentro del Programa SUCCESS. In *Sustainable Coastal Communities and Ecosystems (SUCCESS), a Component of the Global Integrated Management of Coastal and Freshwater Systems (IMCAFS) Program*, ed. USAID, 38. EPP-A-00-04-00014-00: USAID.

Hogarth, P. J. 1999. *The Biology of Mangroves*. New York: Oxford University Press.

———. 2007. *The Biology of Mangroves and Seagrasses*. Oxford: Oxford University Press.

Höhle, J., and M. Höhle. 2009. Accuracy assessment of digital elevation models by means of robust statistical methods. *ISPRS Journal of Photogrammetry and Remote Sensing* 64 (4):398-406.

Holland, M. B., F. de Koning, M. Morales, L. Naughton-Treves, B. E. Robinson, and L. Suárez. 2014. Complex Tenure and Deforestation: Implications for Conservation Incentives in the Ecuadorian Amazon. *World Development* 55:21-36.

Imbert, D., and B. Rollet. 1989. Phytomasse aerienne et production primaire dans la mangrove du Grand Cul-de-Sac Marin (Guadeloupe, Antilles francaises). *Bulletin d'écologie*.

Intergovernmental Panel on Climate Change. 2007. Contribution of Working Group I to the Fourth Assessment Report of the Intergovernmental Panel on Climate Change: The Physical Science Basis. In *Contribution of Working Groups I, II and III to the Fourth Assessment Report of the Intergovernmental Panel on Climate Change*, eds. R. K. Pachauri and A. Reisinger, 104. Cambridge, UK.

Jensen, J. R. 2005. *Introductory Digital Image Processing: A Remote Sensing Perspective*. 3rd Edition ed. Upper Saddle River, NJ: Prentice Hall.

Jensen, J. R. 2007. *Remote Sensing of the Environment: An Earth Resource Perspective*. 2nd Edition ed. Upper Saddle River, NJ: Prentice Hall.

Kauffman, J. B., and D. Donato. 2012. *Protocols for the Measurement, Monitoring and Reporting of Structure, Biomass and Carbon Stocks in Mangrove Forests*. Bogor, Indonesia: Center for International Forestry Research (CIFOR).

Kauffman, J. B., C. Heider, T. G. Cole, K. A. Dwire, and D. C. Donato. 2011. Ecosystem Carbon Stocks of Micronesian Mangrove Forests. *Wetlands* 31 (2):343-352.

Komiyama, A., J. E. Ong, and S. Poungparn. 2008. Allometry, Biomass, and Productivity of Mangrove Forests: A Review. *Aquatic Botany* 89 (2):128-137.

Labastida, E. 1995. Diagnóstico Económico de las Actividades Relacionadas con la Zona de Manglar con énfasis en las Unidades de Producción Camaronera. In *Estudio de las Alternativas de Manejo del Área Comprendida entre los Ríos Cayapas y Mataje, Provincia de Esmeraldas*, ed. E. Inefan, Appendix 9. Quito, Ecuador: EcoCiencia & INEFAN.

Laffoley, D., and G. D. Grimsditch. 2009. *The management of natural coastal carbon sinks*: Iucn.

Lugo, A. E., and C. Patterson-Zucca. 1977. The impact of low temperature stress on mangrove structure and growth. *Tropical Ecology* 18 (2):149-161.

Madsen, J. E., R. Mix, and H. Balslev. 2001. *Flora of Puna Island: Plant Resources on a Neotropical Island*. Langelandsgade, Denmark: Aarhus University Press.

Naughton-Treves, L., and C. Day. 2012. Lessons about Land Tenure, Forest Governance and REDD+ Case Studies from Africa, Asia and Latin America, eds. L. Naughton-Treves and C. Day. Madison: UW-Madison Land Tenure Center. Original edition,

http://www.nelson.wisc.edu/ltc/docs/Lessons-about-Land-Tenure-Forest-Governance-and-REDD.pdf.

Ocampo-Thomason, P. 2006. Mangroves, People and Cockles: Impacts of the Shrimp-Farming Industry on Mangrove Communities in Esmeraldas Province, Ecuador. In *Environment and Livelihoods in Tropical Coastal Zones: Managing Agriculture-Fishery-Aquaculture Conflicts (Comprehensive Assessment of Water Management in Agriculture Series)*, eds. C. T. Hoanh, T. P. Tuong, J. W. Gowing and B. Hardy, 140 - 153. London: Oxford University Press.

Ochoa, E. 1997. Majagual: The Tallest Mangroves in the World. *Intercoastal Network* Special Edition (1):1.

Ong, J. E. 2002. The Hidden Costs of Mangrove Services: Use of Mangroves for Shrimp Aquaculture. In *International Science Roundtable for the Media*. Bali, Indonesia: ICSU, IGBP, IHDP, WCRP, DIVERSITAS, STAR.

Pool, D. J., S. C. Snedaker, and A. E. Lugo. 1977. Structure of Mangrove Forests in Florida, Puerto-Rico, Mexico, and Costa-Rica. *Biotropica* 9 (3):195-212.

Refaeilzadeh, P., L. Tang, and H. Liu. 2009. Cross-validation. In *Encyclopedia of database systems*, 532-538: Springer.

Ricklefs, R. E., and R. E. Latham. 1993. Global patterns of diversity in mangrove floras. *Species diversity in ecological communities: historical and geographical perspectives. University of Chicago Press, Chicago*:215-229.

Rovai, A., P. Riul, R. Twilley, E. Castañeda‐Moya, V. Rivera‐Monroy, A. Williams, M. Simard, M. Cifuentes‐Jara, R. Lewis, and S. Crooks. 2016. Scaling mangrove aboveground biomass from site‐level to continental‐scale. *Global Ecology and Biogeography*.

Ruesch, A., and H. K. Gibbs. 2008. New IPCC Tier-1 Global Biomass Carbon Map For the Year 2000: Carbon Dioxide Information Analysis Center, Oak Ridge National Laboratory, Oak Ridge, Tennessee.

Saenger, P., and S. C. Snedaker. 1993. Pantropical Trends in Mangrove Aboveground Biomass And Annual Litterfall. *Oecologia* 96 (3):293-299.

Siikamäki, J., J. N. Sanchirico, and S. L. Jardine. 2012. Global Economic Potential for Reducing Carbon Dioxide Emissions from Mangrove Loss. *Proceedings of the National Academy of Sciences* (Early Edition).

Smith III, T. J., and K. R. T. Whelan. 2006. Development of allometric relations for three mangrove species in South Florida for use in the Greater Everglades Ecosystem restoration. *Wetlands Ecology and Management* 14 (5):409-419.

Snedaker, S. C. 1986. Traditional Uses of South American Mangrove Resources and the Socio-Economic Effect of Ecosystem Changes. In *Workshop on Man in the Mangroves*, eds. P. Kunstadter, E. C. F. Bird and S. Sabhasri, 102-112. Tokyo, JP: United Nations University.

Soares, M. L. G., and Y. Schaeffer-Novelli. 2005. Above-Ground Biomass of Mangrove Species: Analysis of Models. *Estuarine Coastal and Shelf Science* 65 (1-2):1-18.

Spalding, M., F. Blasco, and C. Field eds. 1997. *World Mangrove Atlas*. Okinawa, JP: International Society for Mangrove Ecosystems.

Spalding, M., M. Kainuma, and L. Collins. 2010. *World Atlas of Mangroves*. London, UK: Earthscan.

Stephenson, N. L., A. J. Das, R. Condit, S. E. Russo, P. J. Baker, N. G. Beckman, D. A. Coomes, E. R. Lines, W. K. Morris, N. Ruger, E. Alvarez, C. Blundo, S. Bunyavejchewin, G. Chuyong, S. J. Davies, A. Duque, C. N. Ewango, O. Flores, J. F. Franklin, H. R. Grau, Z. Hao, M. E. Harmon, S. P. Hubbell, D. Kenfack, Y. Lin, J. R. Makana, A. Malizia, L. R. Malizia, R. J. Pabst, N. Pongpattananurak, S. H. Su, I. F. Sun, S. Tan, D. Thomas, P. J. van Mantgem, X. Wang, S. K. Wiser, and M. A. Zavala. 2014a. Rate of tree carbon accumulation increases continuously with tree size. *Nature* 507 (7490):90-93.

———. 2014b. Rate of Tree Carbon Accumulation Increases Continuously with Tree Size. *Nature* Advance Online Publication.

Tomlinson, P. B. 1986. *The Botany of Mangroves*. New York: Cambridge University Press.

Twilley, R. R., R. H. Chen, and T. Hargis. 1992. Carbon Sinks in Mangroves and Their Implications to Carbon Budget of Tropical Coastal Ecosystems. *Water Air and Soil Pollution* 64 (1-2):265-288.

Wetlands International. 2004. Ramsar Sites Information Service, 1-7: Wetlands.

# Supplemental Material

## Supplementary Introduction.

### Global Mangrove Location

Mangrove forests straddle the tropical and near tropical coastlines and estuaries of Asia, Africa, Europe, Australia, and the Americas with species designated as mangrove present in over 100 countries (Spalding, Kainuma, and Collins 2010; Hamilton and Casey 2016). Although generally confined to the tropics mangroves can actually be found as far north as 32.3502° within Walsingham Nature Reserve, Bermuda and as far south as -38.7621° on Snake Island, Australia (Spalding, Kainuma, and Collins 2010; Loftis et al. 2014; Hamilton and Casey 2016). As mangroves generally require a minimum water temperature of approximately 19 °C (Waisel 1972) or 20 °C (Ellison, Farnsworth, and Merkt 1999) and an average monthly air temperature greater than 20 °C (Chapman 1976), these maximum latitudinal ranges are likely supported by the warm ocean currents of the Gulf Stream and the southern flow of the East Australian current.

## Supplementary Methods.

### Biomass to carbon uncertainty

Our biomass to carbon ratio of 46.4 percent is a conservative estimate that with the range of estimates that typically average slightly below 50 percent (Kauffman and Donato 2012). Twilley, Chen, and Hargis (1992) use an estimate of 45 percent carbon per unit of mangrove biomass but report no error statistics or source for this estimate. Kauffman et al. (2011) and Kauffman and Donato (2012) use a carbon conversion of 46.4 percent across multiple species and estimate a common practice range of between 1:0.46 and 1:0.5 but do not report an uncertainty measure.

**Supplementary Discussion.**

Extracting the Cayapas-Mataje and Chone estuarine carbon data from the IPCC-compliant 1 km grids (Ruesch and Gibbs 2008) reveals two potential reasons for the differences between the carbon measures. Upon examination of the IPCC compliant grid methodology, the land cover classification system utilized does not contain a distinct mangrove forest category and this may contribute to the underestimation of forest carbon in these estuaries as mangrove likely contain more carbon that other forest type. Furthermore, the IPCC-compliant grids are based on a land mask that may also result in the omission of some aquatic forest systems. By comparing the mangrove carbon findings presented with data from Hamilton and Lovette (2015), which match the IPCC-compliant carbon values to within 1 percent, it can be demonstrated that the factor contributing to the mangrove carbon differences between this study and earlier study and the IPCC-compliant grids is the field measurement component of this study.

It is likely that global mangrove estimates, upscaled from local field research of this nature, will continue to improve the accuracy and representation of regional mangrove carbon holdings and their $CO_2$ sequestration potential. Whether the regionally important finding presented in this article will have a meaningful impact on the modeled global mangrove carbon estimates is likely based on the wider applicability of these findings to other mangroves in neighboring countries and the wider region. Further research is required to determine if these findings are replicated in Ecuador's southern estuaries and across the northern border into Colombia, or if the *Rhizophora mangle* carbon storage values exist only in this region.

**Supplemental Figure 1**

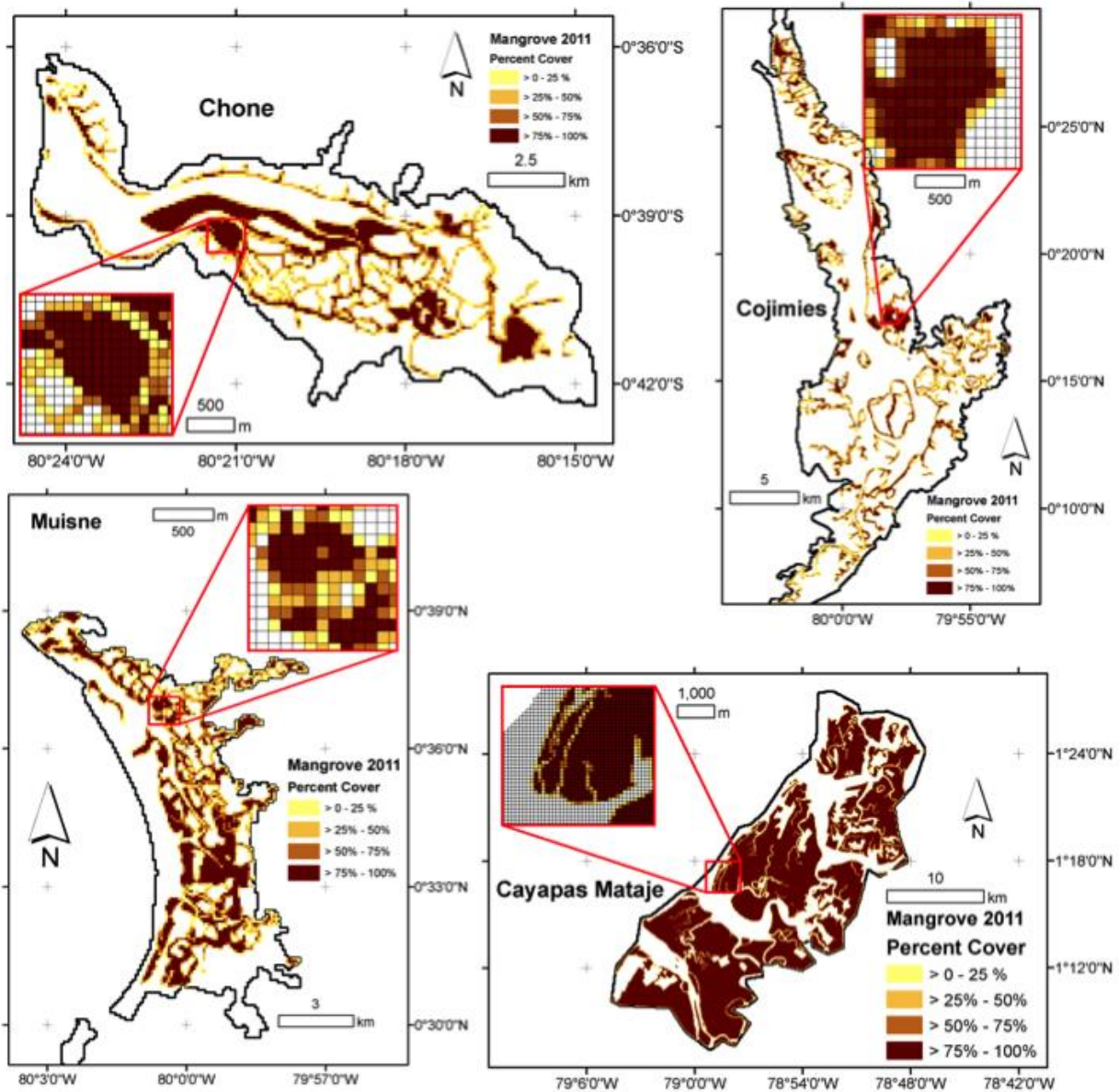

The mangrove coverage percent for each 1 ha grid cell is represented for each of the four estuaries. Artifacts of the aquaculture can be seen in some of the estuaries, leaving only fringe areas of mangrove forest.

**Supplemental Figure 2**

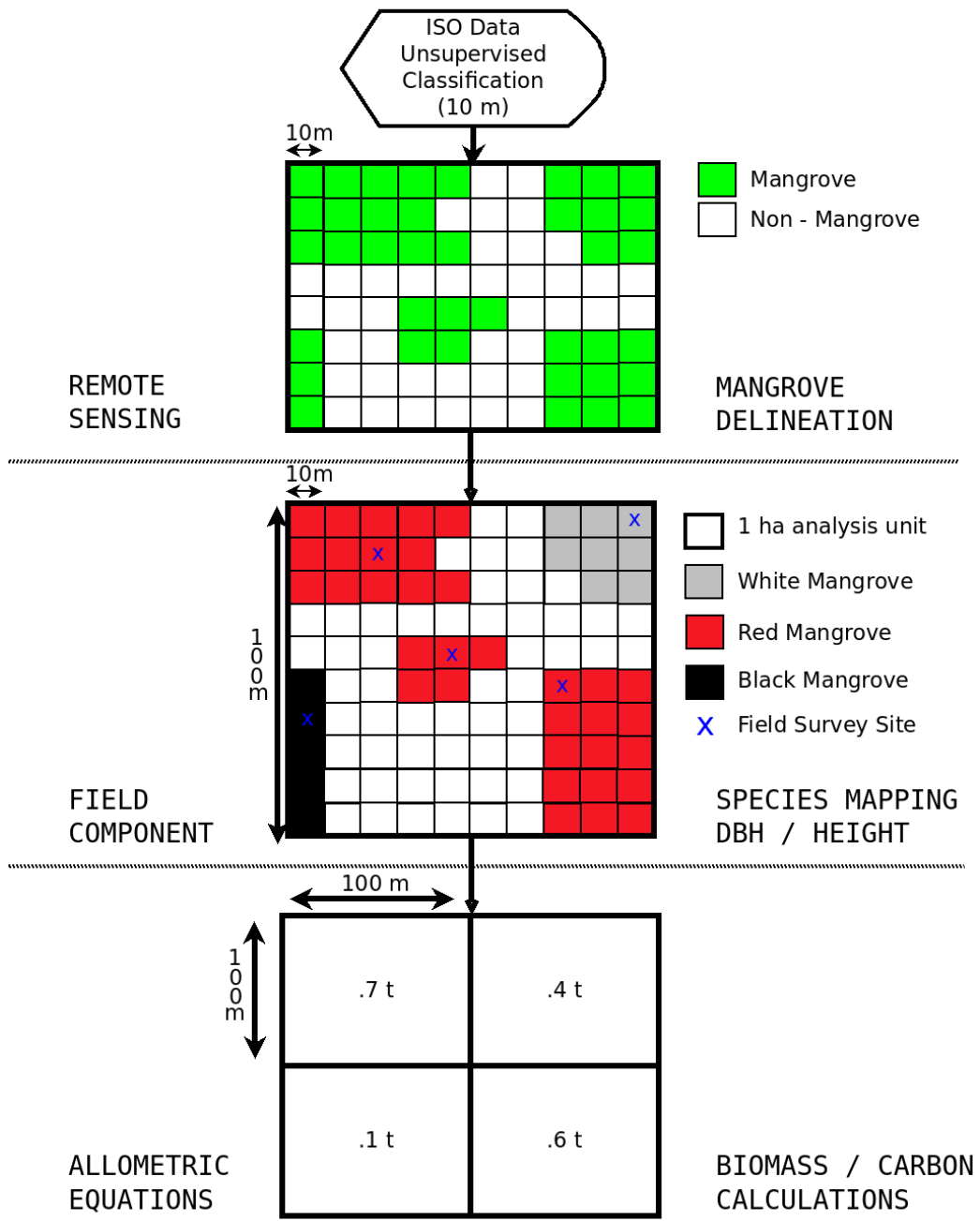

A schematic of the process that converts mangrove tree cover at the 10 m level to carbon at the 1 ha level. Panel 1 represent the 10 m cells designated as mangrove by the remote sensing process. Panel 2 show how these 10 m cells are converted to species stands and the location of the hypothetical field surveys. Panel 3 shows how the many 10 m cells are converted into 1 ha cells with carbon estimates.

**Supplemental Figure 3**

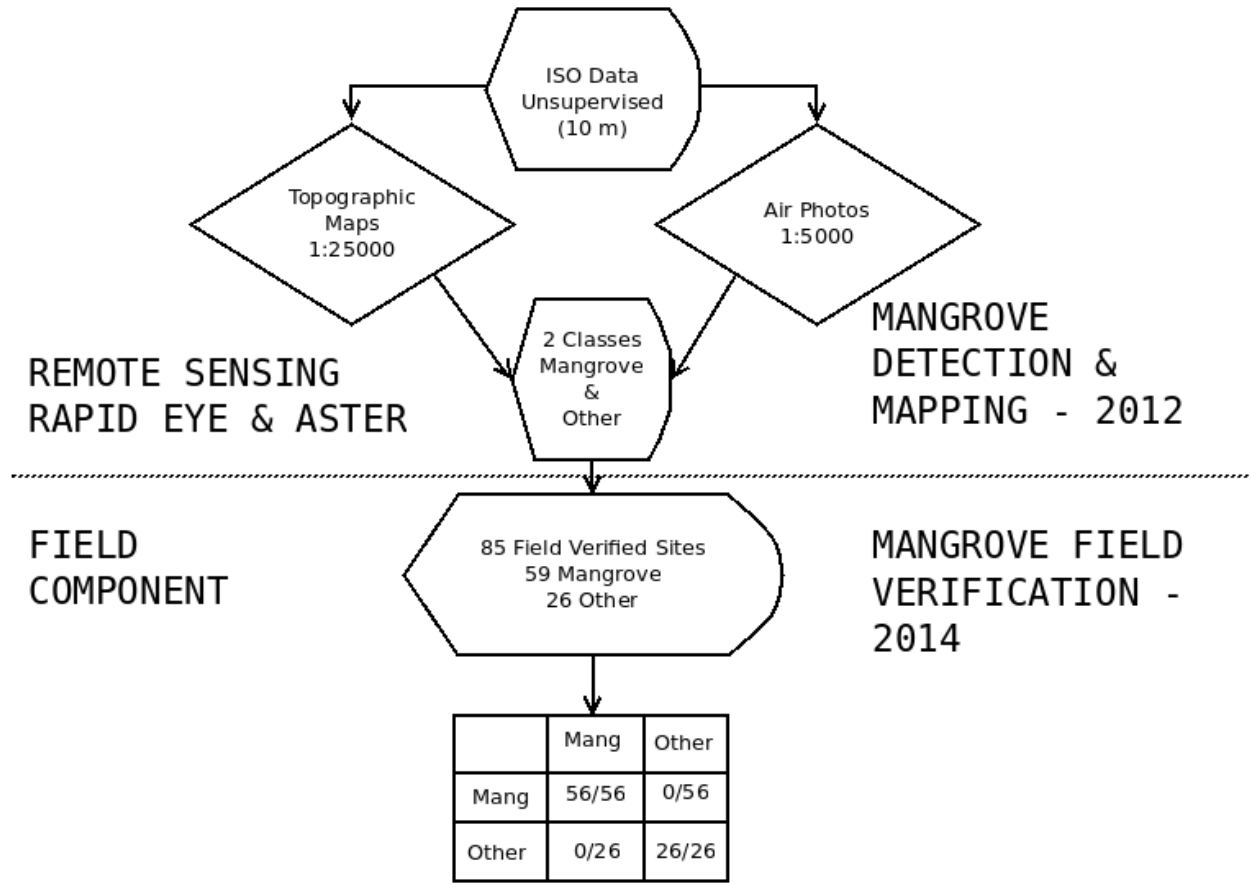

A schematic of the process of mangrove verification process for land cover.

# Supplemental Table 1

| Study | Location | Species | Equation | Per tree AGMB g/tree 10 cm | Per tree AGMB g/tree 20 cm | Per tree AGMB t/tree 10 cm | Per tree AGMB t/tree 20 cm | AGMB**** t/ha 10 cm | AGMB**** t/ha 20 cm | BGMB t/ha 10 cm | BGMB t/ha 20 cm | CMB t/ha 10 cm | CMB t/ha 20 cm | Aboveground Carbon t/ha 10 cm | Aboveground Carbon t/ha 20 cm | Combined Carbon t/ha 10 cm | Combined Carbon t/ha 20 cm |
|---|---|---|---|---|---|---|---|---|---|---|---|---|---|---|---|---|---|
| Soares and Schaeffer-Novelli (2005) | Brazil | Rhizophora mangle | e^(4.89219 + 2.61724 * ln(DBH)) | 55,193.90 | 338,656.07 | 0.06 | 0.34 | 44.16 | 270.92 | 22.96 | 140.88 | 67.12 | 411.81 | 20.49 | 125.71 | 31.14 | 191.08 |
| Fromard et al. (1998) | French Guiana | Rhizophora mangle | 128.2 * (DBH)^2.6 | 51,037.34 | 309,432.56 | 0.05 | 0.31 | 40.83 | 247.55 | 21.23 | 128.72 | 62.06 | 376.27 | 18.95 | 114.86 | 28.80 | 174.59 |
| Fromard et al. (1998) | French Guiana | Laguncularia racemosa | 102.3 * (DBH)^2.5 | 32,350.10 | 182,999.80 | 0.03 | 0.18 | 25.88 | 146.40 | 13.46 | 76.13 | 39.34 | 222.53 | 12.01 | 67.93 | 18.25 | 103.25 |
| Fromard et al. (1998)*** | French Guiana | Avicennia germinans | 200.4 * (DBH)^2.1 | 25,228.87 | 108,158.51 | 0.03 | 0.11 | 20.18 | 86.53 | 10.50 | 44.99 | 30.68 | 131.52 | 9.36 | 40.15 | 14.23 | 61.03 |
| Fromard et al. (1998) | French Guiana | Avicennia germinans | 140 * (DBH)^2.4 | 35,166.41 | 185,609.43 | 0.04 | 0.19 | 28.13 | 148.49 | 14.63 | 77.21 | 42.76 | 225.70 | 13.05 | 68.90 | 19.84 | 104.73 |
| Smith III and Whelan (2006) | USA | Rhizophora mangle | 772.7 * (DBH)^1.731 | 41,592.11 | 138,068.25 | 0.04 | 0.14 | 33.27 | 110.45 | 17.30 | 57.44 | 50.58 | 167.89 | 15.44 | 51.25 | 23.47 | 77.90 |
| Smith III and Whelan (2006) | USA | Laguncularia racemosa | 362.2 * (DBH)^1.93 | 30,828.22 | 117,472.53 | 0.03 | 0.12 | 24.66 | 93.98 | 12.82 | 48.87 | 37.49 | 142.85 | 11.44 | 43.61 | 17.39 | 66.28 |
| Smith III and Whelan (2006) | USA | Avicennia germinans | 402.7 * (DBH)^1.934 | 34,592.47 | 132,182.40 | 0.03 | 0.13 | 27.67 | 105.75 | 14.39 | 54.99 | 42.06 | 160.73 | 12.84 | 49.07 | 19.52 | 74.58 |
| Imbert and Rollet (1989) | Guadeloupe | Rhizophora mangle | 177.9 * (DBH)^2.4176 | 46,534.60 | 248,625.73 | 0.05 | 0.25 | 37.23 | 198.90 | 19.36 | 103.43 | 56.59 | 302.33 | 17.27 | 92.29 | 26.26 | 140.28 |
| Imbert and Rollet (1989) | Guadeloupe | Laguncularia racemosa | 54.2 * (DBH)^2.5367 | 32,415.36 | 188,093.46 | 0.03 | 0.19 | 25.93 | 150.47 | 13.48 | 78.25 | 39.42 | 228.72 | 12.03 | 69.82 | 18.29 | 106.13 |
| Imbert and Rollet (1989) | Guadeloupe | Avicennia germinans | 208.8 * (DBH)^2.239 | 36,201.83 | 170,897.87 | 0.04 | 0.17 | 28.96 | 136.72 | 15.06 | 71.09 | 44.02 | 207.81 | 13.44 | 63.44 | 20.43 | 96.42 |

*** only valid for DBH < 4cm
**** 800 trees / ha

## Supplementary References

Hamilton, S. E., and J. Lovette. 2015. Ecuador's mangrove forest carbon stocks: a spatiotemporal analysis of living carbon holdings and their depletion since the advent of commercial aquaculture. PLoS ONE 10 (3):e0118880.

Kauffman, J. B., and D. Donato. 2012. Protocols for the Measurement, Monitoring and Reporting of Structure, Biomass and Carbon Stocks in Mangrove Forests. Bogor, Indonesia: Center for International Forestry Research (CIFOR).

Kauffman, J. B., C. Heider, T. G. Cole, K. A. Dwire, and D. C. Donato. 2011. Ecosystem Carbon Stocks of Micronesian Mangrove Forests. Wetlands 31 (2):343-352.

Ruesch, A., and H. K. Gibbs. 2008. New IPCC Tier-1 Global Biomass Carbon Map For the Year 2000: Carbon Dioxide Information Analysis Center , Oak Ridge National Laboratory, Oak Ridge, Tennessee.

Twilley, R. R., R. H. Chen, and T. Hargis. 1992. Carbon Sinks in Mangroves and Their Implications to Carbon Budget of Tropical Coastal Ecosystems. Water Air and Soil Pollution 64 (1-2):265-288.